\documentclass[prl,superscriptaddress,showpacs,amssymb,amsmath,
amsfonts,aps,twocolumn,secnumarabic,floatfix]{revtex4}
\usepackage{graphics}
\usepackage{epsfig}
\begin{document} 
\bibliographystyle{try} 

\topmargin -0.9cm 
 
 \title{Unitarity constraints on charged pion photoproduction at large $p_{\perp}$}

\newcommand*{\JLAB }{ Thomas Jefferson National Accelerator Facility, Newport News, Virginia 23606} 
\affiliation{\JLAB } 

\author{J.M.~Laget}
     \affiliation{\JLAB}

\date{\today} 

\begin{abstract} 

Around $\theta_{\pi}=$~90$^\circ$, the coupling to the $\rho^\circ N$ channel leads to a good accounting of the charged pion exclusive photoproduction cross section in the energy range 3~$<E_{\gamma}<$~10~GeV, where experimental data exist.  Starting from a Regge Pole approach that successfully describes vector meson production,  the singular part of the corresponding box diagrams (where the intermediate vector meson-baryon pair propagates on-shell) is evaluated without any further assumptions (unitarity). Such a treatment provides an explanation of the $s^{-7}$ scaling of the cross section. Elastic rescattering of the charged pion improves the basic Regge pole model at forward and backward angles.

\end{abstract} 
 
 
\maketitle

At high photon energies (let say $E_{\gamma}>$3~GeV) and forward angles, charged pion photoproduction on nucleon is well accounted for by the exchange, in the $t$-channel, of the $\pi$ and the $\rho$ Regge linear trajectories~\cite{Gui97a,Gui97b}. At backward angles,  the exchange of the nucleon and the Delta linear Regge trajectories,  in the {u}-channel, leads also to a good account of the cross section~\cite{Gui97a,Gui98}. Around $\theta_{\pi}=$~90$^\circ$ (large $p_{\perp}$), this basic Regge model misses the data by orders of magnitude, and we have advocated the use of saturating trajectories to fill in this gap~\cite{Gui97a,Gui97b}. This was a poor man way to incorporate quark degrees of freedom and recover the $s^{-7}$ scaling behavior of the experimental cross section at  $\theta_{\pi}=$~90$^\circ$ (being $\sqrt s=W$ the total c.m. energy). Such a scaling was considered as the evidence of quark degrees of freedom~\cite{Bro80}, however any attempt to compute the cross section of this channel, within pertubative QCD, failed in describing the data~\cite{Fa91,Kro00,Kro04}.

This note proposes a more natural explanation in terms of channel couplings. Unitarity tells us that  the amplitude of an exclusive reaction is driven by the overlap of the production and absorption  amplitudes of all the possible intermediate states. The integral runs over the angles of these intermediate states. At low energies, the production and absorption  amplitudes are more or less flat, and many intermediate states may contribute. A full coupled channel treatment is mandatory, and this is the duty of the Excited Baryon Center (EBAC) at Jefferson Laboratory. At high energies, on the contrary, the angular distributions are strongly forward peaked, and the unitary integral picks only the few intermediate states that have the highest production cross section. In the 3~$<E_{\gamma}<$~10~GeV energy range, the $N(\gamma,\rho)N$ channel overwhelms the others; typically its cross section is more than ten times the cross section of the $\omega$  or the $\pi$ production channels. Among those less important channels, a special attention must be paid however to the pion elastic rescattering: Although the pion production is the same in the pole  and in the resctattering amplitude, the pion elastic scattering amplitude is almost purely absorptive in this energy range. The consequence is that the corresponding unitary integral interferes destructively with the  pole amplitude at the most forward and backward angles.   

\begin{figure}[hbt]
\begin{center}
\epsfig{file=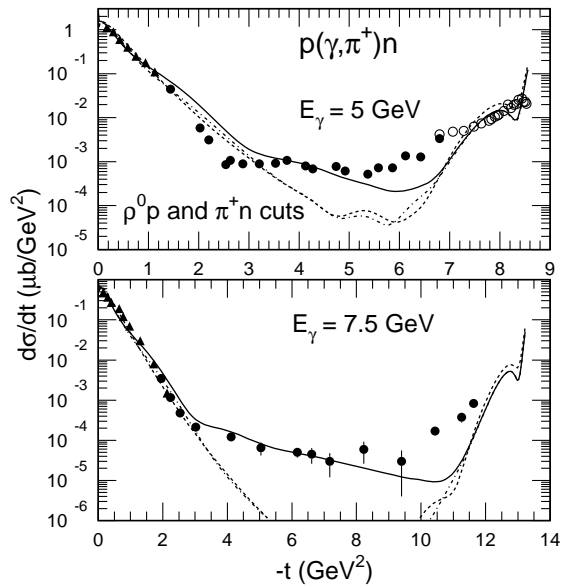,width=3.in}
\caption[]{The cross section of the $p(\gamma,\pi^+)n$ reaction at $E_{\gamma}=$ 5~GeV and 7.5~GeV. The dashed line curves are the basic Regge Pole model. The dash-dotted line curves take into account the contribution of the $\pi^+n$ elastic cut. The full line curves take also into account the contribution of the inelastic $\rho^{\circ}p$ cut. The references to the experimental data can be found in refs.~\cite{Gui97a,Gui97b}.}
\label{dsdt_piplus}
\end{center}
\end{figure}

Figure~\ref{dsdt_piplus} summarizes these findings, for the $p(\gamma,\pi^+)n$. reaction. The contribution of the $\rho^{\circ}p$ cut alone reproduces the angular and energy variations of the experimental data around $\theta_{\pi}=$~90$^\circ$. The $\pi^+n$ elastic cut brings down $u-$channel Regge pole contribution close to the experimental data at backward angles, but affects little the forward angle cross section.

The basic Regge model is fully described in ref.~\cite{Gui97b}. Instead of their expression in terms of $\gamma$ matrices, I use the expression of the  $\pi$ and $\rho$ $t$-channel exchange amplitudes in terms of $\sigma$ matrices that are given in the appendix of ref.~\cite{La06}. The expression of the nucleon and Delta $u$-channel exchange amplitudes has been given in the Guidal's thesisÃ~\cite{Gui98}, but not published elsewhere. For the sake of completeness, I give the reduction in terms of $\sigma$ matrices that I use.

The nucleon exchange amplitude takes the form:
\begin{eqnarray}  
T_N&=& -i\; e \mu_n g_{\pi}  \frac{\sqrt{(E_i+m)(E_f+m)}}{2m} {\cal P}_N^R(u)
\nonumber \\
&& \left( \lambda_f \left |
\vec{\sigma} \cdot \vec{k_{\gamma}} \times \vec{\epsilon}
 \;\;\vec{\sigma} \cdot \left[ 
 \frac{\vec{k_{\pi}}-\vec{p_i}}{\sqrt{m^2+ (\vec{k_{\pi}}-\vec{p_i})^2}+m} 
\right .\right .\right .    \nonumber \\ &&
 \left . \left . \left . + \frac{\vec{p_i}}{E_i+m} \right]
\right | \lambda_i \right )
\label{uN}
\end{eqnarray}
Where $(E_i, \vec{p_i})$ and $(E_f, \vec{p_f})$ are the four momenta of the target proton and the final neutron, respectively. Where $\vec{k_{\gamma}}$ is the momentum of the ingoing photon and $\vec{\epsilon}$ is its polarization. Where $(E_{\pi}, \vec{k_{\pi}})$ is the four momentum of the outgoing pion. The four momentum transfer in the $u$-channel is $u= (k_{\pi}-p_i)^2$. The magnetic moment of the neutron is $\mu_n=$ -1.91, and the pion nucleon coupling constant is $g^2_{\pi}/4\pi=$~14.5.

To be consistent with the backward angle cross section of the $p(\gamma,\omega)p$ channel~\cite{Gui98,La00,Ca02}, where only the proton can be exchanged in the $u$-channel, I use the non degenerated Regge propagator
\begin{eqnarray}  
{\cal P}_N^R&=& \left (\frac{s}{s_{\circ}} \right)^{\alpha_N -0.5}
 \alpha'_N \Gamma (0.5-\alpha_N)
\frac{1- e^{-i\pi(\alpha_N+0.5)}}{2}
 \label{prop_N}
\end{eqnarray}
where $s_{\circ}=$ 1~GeV$^2$ and where the nucleon trajectory is
$  \alpha_N= -0.37 +\alpha'_N \; u$,
with $\alpha'_N=$~0.98.

Since the Delta exchange amplitude contributes little to the cross section I do not give its expression here.

These $u$-channel contributions overestimate, by about a factor two, the backward angle cross section. In refs.~\cite{Gui97a,Gui98} we have renormalized  those amplitudes, on the basis that the physical region is far from the nucleon  or the Delta pole. I do not  use such a reduction form factor, since the elastic $\pi^+n$ cut naturally brings down the Regge cross section close to experiment.

\begin{figure}[hbt]
\begin{center}
\hspace{-2.2cm}
\epsfig{file=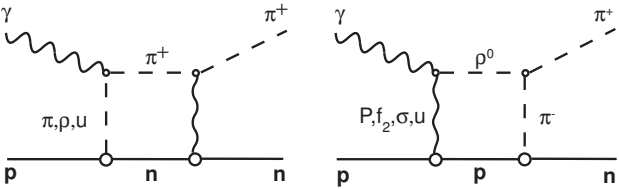,width=4.in}
\caption[]{The relevant graphs in the $\gamma p \rightarrow \pi^+ n$ reaction. Left: Elastic $\pi^+ n$ cut. Right: $\rho$-nucleon unitary cut.}
\label{graph_cut}
\end{center}
\end{figure}

Fig.~\ref{graph_cut} depicts the two cuts that are dominant in this study. Only the $\pi^+$ production channel is drawn, and the corresponding amplitudes will be given. The $\pi^-$ production amplitudes are trivially related.

The second graph in Fig.~\ref{graph_cut} depicts the production of the $\rho$ meson followed by the reabsorption of one of its decay pions by the nucleon. 
The corresponding rescattering amplitude takes the form:
\begin{eqnarray}  
T_{\rho N}&=& \int \frac{d^3\vec{p}}{(2\pi)^3}
\frac{m}{E_p}
 \frac{1}{P^2_{\rho}-m^2_{\rho}+i\epsilon}
T_{\gamma p\rightarrow \rho^{0}p}T_{\rho^{0} p \rightarrow \pi^{+} n}
\end{eqnarray}
where the integral runs over the three momentum $\vec{p}$ of the intermediate nucleon, of which the mass is $m$ and the energy is $E_p=\sqrt{p^2+m^2}$. The four momentum  and the mass of the intermediate $\rho$ are respectively $P_{\rho}$ and $m_{\rho}$. The integral can be split into a singular part, that involves on-shell matrix elements, and a principal part~$\cal{P}$:
\begin{eqnarray}
 T_{\rho N}&=& -i\frac{p_{c.m.}}{16\pi^2}\frac{m}{\sqrt{s}}
 \int  d{\Omega} \left[T_{\gamma p\rightarrow \rho^{0} p}(t_{\gamma})
 T_{\rho^{0} p \rightarrow \pi^{+} n}(t_{\pi})
 \right]
 \nonumber \\ && 
 +\cal{P}
 \label{sing}
\end{eqnarray}
where $p_{c.m.}=\sqrt{(s-(m_{\rho}-m)^2)(s-(m_{\rho}+m)^2)/4s}$  is the on-shell momentum  of the intermediate proton, for  the c.m. energy $\sqrt s$. The two fold integral runs over the solid angle $\Omega$ of the intermediate proton. The four momentum transfer between the incoming photon and the $\rho$ is $t_{\gamma}=(k_{\gamma}-P_{\rho})^2$, while the four momentum transfer between the $\rho$ and the outgoing pion is $t_{\pi}=(k_{\pi}-P_{\rho})^2$. The summation over all the spin indices of the intermediate particles is meant. 

I neglect the principal part. The singular part of the integral relies entirely on on-shell matrix elements and is parameter free as long as one has a good description of the production and the absorption processes

For photo-production of vector mesons, $T_{\gamma p\rightarrow \rho^{0} p}$, I use the Regge model~\cite{La00,Ca02} which reproduces the world set of data in the entire angular range (see for instance Fig.~3 in ref.~\cite{Ba01}). It has been extended to electro-production~\cite{Ca03,La04} and also reproduces the data~\cite{CyXX,Gui07}. The model takes into account the exchange of the Pomeron, the $f_2$ and $\sigma$ mesons in the $t$-channel, as well as the exchange of the nucleon and the Delta in the $u$-channel.

\begin{figure}[hbt]
\begin{center}
\epsfig{file=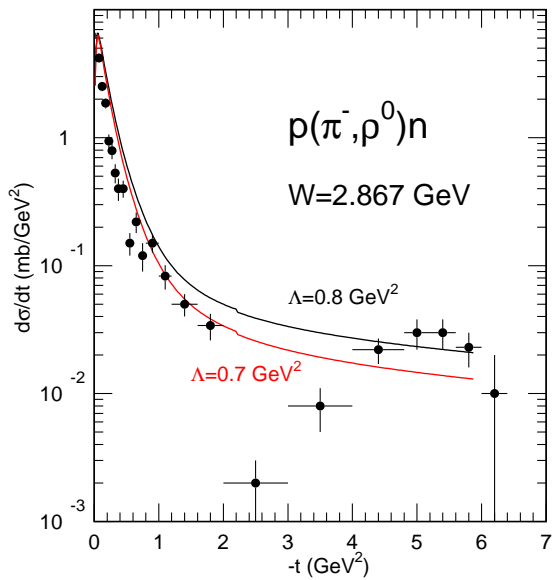,width=3.in}
\caption[]{(Color online) The cross section of the reaction $p(\pi^-,\rho^{\circ})n$. The experimental data are from ref.~\cite{PiroXX} }
\label{rho_pi}
\end{center}
\end{figure}

For the reabsorption of the $\rho$ meson, $T_{\rho^{0} p \rightarrow \pi^{+} n}$, I use a pion exchange Regge description which nicely reproduces the data in the energy range that is considered in this study (Figure~\ref{rho_pi}). The corresponding matrix element takes the form:
\begin{eqnarray}
T_{\rho\pi}&=& \sqrt{2} g_{\pi} g_{\rho} \frac{\sqrt{(E_p+m)(E_f+m)}}{2m}
{\cal P}_{\pi}^{R}(t_{\pi}) F_1(t_{\pi})
 \nonumber \\ && 
\left[ \vec{\epsilon_{\rho}} \cdot \left( \vec{k_{\pi^+}} -\vec{p}+\vec{p_f}\right)
-\epsilon^{\circ}_{\rho} \left( E_{\pi^+}-E_p+E_f \right)  \right]
 \nonumber \\ && 
\left( \lambda_f \left|\vec{\sigma} \cdot 
\left[  \frac{\vec{p}}{E_p+m} -\frac{\vec{p_f}}{E_f+m}  \right]
 \right| \lambda \right)
 \nonumber \\ && 
 \label{rho_to_pi}
\end{eqnarray}
where $(\epsilon^{\circ},\vec{\epsilon})$ is the intermediate $\rho$ polarization, and where $(E_f,\vec{p}_f)$ and $(E_{\pi^+},\vec{k}_{\pi^+})$ are the four momenta of  the  outgoing neutron and pion respectively. The pion nucleon coupling constant is $g^2_{\pi}/4\pi=$~14.5, and the rho decay constant is $g^2_{\rho}/4\pi=$~5.71.  ${\cal P}_{\pi}^{R}$ is the pion Regge propagator, with a degenerated non rotating saturating trajectory, and $ F_1$ is the nucleon form factor. I use the same expressions and couplings as in ref.~\cite{Gui97b}, except for the cut-off mass in  $ F_1$ that I take as $\Lambda=$ 0.7~GeV$^2$. For $\pi^-$ production, the amplitude is multiplied by $-1$, for symmetry and charge reasons.

The first graph in Fig.~\ref{graph_cut} depicts the elastic rescattering of the pion. The singular part of the corresponding rescattering amplitude takes the form:
\begin{eqnarray}
 T_{\pi N}&=& -i\frac{p'_{c.m.}}{16\pi^2}\frac{m}{\sqrt{s}}
 \int  d{\Omega} \left[T_{\gamma p\rightarrow \pi^+ n}(t_{\gamma})
 T_{\pi^{+} n \rightarrow \pi^{+} n}(t_{\pi})
 \right]
 \nonumber \\ && 
 \label{pi_el}
\end{eqnarray}
where $p'_{c.m.}=\sqrt{(s-(m_{\pi}-m)^2)(s-(m_{\pi}+m)^2)/4s}$  is the on-shell momentum  of the intermediate proton, for  the c.m. energy $\sqrt s$. The two fold integral runs over the solid angle $\Omega$ of the intermediate proton. The four momentum transfer between the incoming photon and the intermediate $\pi$ is $t_{\gamma}=(k_{\gamma}-P_{\pi})^2$, while the four momentum transfer between the intermediate $\pi$ and the outgoing pion is $t_{\pi}=(k_{\pi}-P_{\pi})^2$. The summation over all the spin indices of the intermediate particles is meant.

The pion photo-production amplitude is the same as in the pole term. I choose a purely absorptive pion nucleon elastic scattering amplitude:
 \begin{eqnarray}
T_{\pi^{+} n \rightarrow \pi^{+} n}=-\frac{\sqrt{s}\; p'_{c.m.}}{m} 
(\epsilon_{\pi} + i)\sigma_{\pi^-p} \exp[\frac{\beta_{\pi}}{2}t_{\pi}]
\label{scat_pin}
\end{eqnarray}
Above $\sqrt{s}\sim 2$~GeV, the total cross section stays constant at the value $\sigma_{\pi^-p}= 30$~mb~\cite{PDG}, and the fit of the differential cross section at forward angles leads to a slope parameter $\beta_{\pi}=6$~GeV$^{-2}$~\cite{La72}. At high energy the ratio between the real and imaginary part of the amplitude is small~\cite{PDG} and I set $\epsilon_{\pi}=0$ in this study.

Under those assumptions, the sum of the Regge pole amplitude and the elastic $\pi N$ cut takes the form:
 \begin{eqnarray}
 T&=& T_{\gamma p\rightarrow \pi^+ n}(t)
 \nonumber \\ 
 && -\frac{p'\,^2_{c.m.}}{16\pi^2}\sigma_{\pi^-p}
 \int  d{\Omega} T_{\gamma p\rightarrow \pi^+ n}(t_{\gamma})
 \exp[\frac{\beta_{\pi}}{2}t_{\pi}]
\label{scat_sum}
\end{eqnarray}
where $t=(k_{\pi}-k_{\gamma})^2$ is the overall four momentum transfer. It clearly shows the purely destructive interference, that is expected for an absorptive rescattering. It compensates the contribution of the $u$-channel Regge poles at the very backward angles. The effect is less important at the very forward angles, simply because the $t$-channel Regge contribution is more than an order of magnitude larger than the $u$-channel one.

\begin{figure}[hbt]
\begin{center}
\epsfig{file=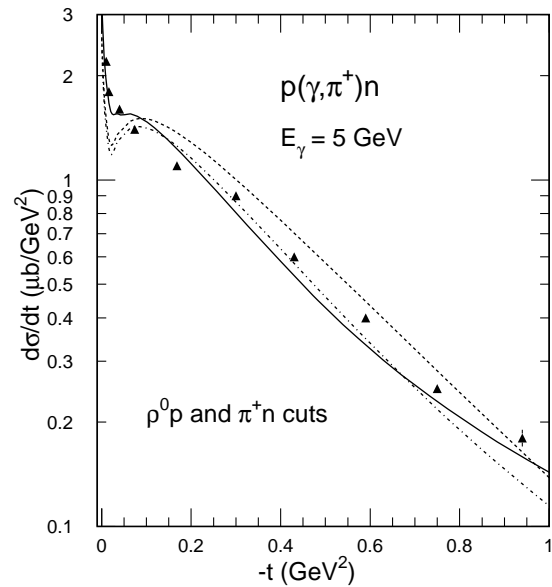,width=3.in}
\caption[]{The cross section of the $p(\gamma,\pi^+)n$ reaction at forward angles. Basic Regge pole model: dotted lines. Elastic pion rescattering cut included: dash-dotted lines. Inelastic $\rho^{\circ}N$ also included: full lines.  Filled triangle: ref~\cite{AnXY}}.
\label{zoom}
\end{center}
\end{figure}

Figure~\ref{zoom} shows how the delicate interference between the pole terms and the cuts reproduces the $\pi^+$ cross section at forward angles. The rise of the cross section at the very forward angle comes the interference between the $t$-channel $\pi$ and $\rho$ exchange and the $s$-channel nucleon exchange that is necessary to restore gauge invariance in the basic Regge model (see ref.~\cite{Gui97b} for an extensive discussion). The $\pi N$ elastic cut brings the model very close to the data at moderate $-t$, while the $\rho N$ inelastic cut fills in the minimum around $-t=$ 0.02~GeV.

Contrary to the earliest Regge approaches, in which the residues of the poles and the effective residues of the cuts were fitted to experiments, the model that I use in this note takes into account the full spin-momentum structure  of the elementary amplitudes and is basically parameter free. As explained in ref.~\cite{Gui97b} the expression of the  Regge pole amplitudes follows the Lagrangian of each vertex, where the coupling constants are determined from the analysis of other independent channels, and the Feymnan propagator is simply replaced by the Regge propagator. The cut amplitudes rely on on-shell elementary amplitudes in the initial state and final state. Again the full spin-momentum dependency is taken into account, and the integral is performed numerically. The results are therefore founded on solid grounds, at least in the domain of validity of the Regge approach, let say above $E_{\gamma}\sim$ 4~GeV ({\it i.e.} above $W\sim$ 3~GeV).

\begin{figure}[hbt]
\begin{center}
\epsfig{file=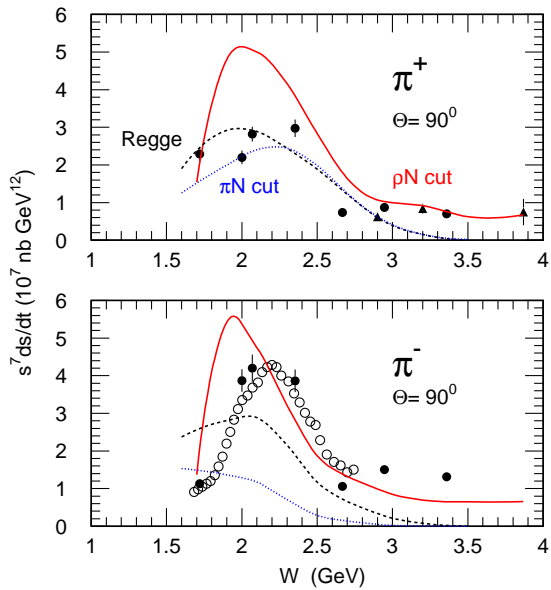,width=3.in}
\caption[]{(Color online) The scaled cross sections of the $p(\gamma,\pi^+)n$ (top) and the $n(\gamma,\pi^-)p$ (bottom) reactions at $\theta_{\pi}=$ 90$^{\circ}$. Basic Regge pole model: (black) dashed lines. Elastic pion rescattering cut included: (blue)dotted lines. Inelastic $\rho^{\circ}N$ also included: (red) full lines. Filled circles: ref~\cite{Zhu05}. Empty circle: ref~\cite{Chen09}. Filled triangle: ref~\cite{AnXY}}.
\label{s7dsdt_90}
\end{center}
\end{figure}

It is interesting to see how this model behaves at lower energies where data have been recently recorded at JLab~\cite{Zhu05,Chen09}. Figure~\ref{s7dsdt_90} shows  the $90^{\circ}$ cross section, scaled by $s^7$, for $\pi^+$ and $\pi^-$ photo-production. In the $\pi^+$ production sector, the model reproduces very well the magnitude and the $s^{-7}$ scaling behavior of the cross section above $W=$ 3~GeV. Below it follows the rise of cross section and goes down at the $\rho$ production threshold. Here, the sharp drop comes from the fact that I have assumed a stable intermediate $\rho$ in the cut: taking into account its width would certainly reduce the height of the theoretical bump and smear its low energy side. Also the principal part of the integral in eq.~\ref{sing} survives below the threshold. Finally, others channels (namely the $\pi \Delta$ channels) may contribute below $W=$ 2~GeV and  one enters into the resonance region. 

The same pattern occurs  in the $\pi^-$ production channel, where the model gives a good account of the data down to $W=$ 2~GeV. One notes that the interference between the basic Regge pole amplitude and the $\pi N$ elastic cut is more important in the $\pi^-$ than in the $\pi^+$ channel. The reason is that the Regge propagators~\cite{Gui97b} have a constant phase in the $\pi^-$ channel, but a rotating phase in the $\pi^+$ channel. The rescattering integral, eq.~\ref{pi_el}, destroys the coherence between the pole terms and the elastic cut in the $\pi^+$ channel, but leaves it intact (eq.~\ref{scat_sum}) in the $\pi^-$ channel.  

\begin{figure}[hbt]
\begin{center}
\epsfig{file=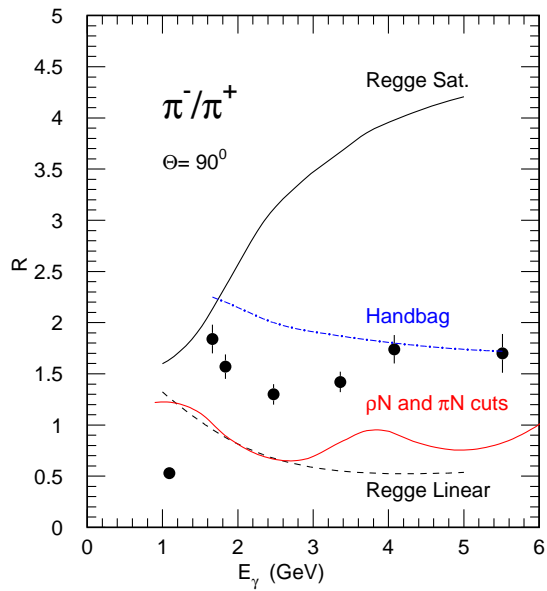,width=3.in}
\caption[]{(Color online) The ratio of the cross sections of the $n(\gamma,\pi^-)p$  and the $p(\gamma,\pi^+)n$  reactions at $\theta_{\pi}=$ 90$^{\circ}$. The data are from ref~\cite{Zhu05}}
\label{R_90}
\end{center}
\end{figure}

Figure~\ref{R_90} shows the ratio of the $\pi^-$ and the $\pi^+$ cross section at $\theta_{\pi}=$ 90$^{\circ}$. The agreement between the data and the basic Regge model~\cite{Gui97b} or the handbag quark based model~\cite{Kro04} (as quoted in~\cite{Zhu05}) is not relevant, since both these models miss the cross section. Among the two models that reproduce the magnitude of the cross sections, the data prefer the coupled channel approach, that I have presented here, rather than the saturating trajectory version of the basic Regge model.  

In summary, the coupling to the $\rho^{\circ}$ meson production channel provides a natural explanation, which does not rely on pertubative QCD, of the magnitude and the scaling with energy  of the cross section of charged pion production at intermediate angles (around $90^{\circ}$). At forward and backward angles the interference with elastic pion rescattering improves the basic Regge pole approach. This model provides us with a good baseline for the analysis of experiments in the range 4$<E_{\gamma}<$ 11~GeV that will become possible with the 12 GeV upgrade of CEBAF at JLab. Below, it reproduces the trend of the deviations from scaling that was recently observed at JLab above the rho meson production threshold.

I acknowledge the warm hospitality at JLab where this work was completed. Jefferson Science Associates operate Thomas Jefferson National Facility for the United States Department of Energy under contract DE-AC05-06OR23177.

\end{document}